\documentclass[prd,tighten,twocolumn,aps]{revtex4}
\usepackage[T1]{fontenc}

\newcommand{\be}{\begin{equation}}
\newcommand{\ee}{\end{equation}}

\begin{document}

\title{The  Extended   ADS/CFT Correspondence\vspace{3mm}\\
{\normalsize 100 years of relativity\\
International Conference on Classical and
Quantum Aspects of Gravity and Cosmology,\\
MASP, S\~ao  Paulo   August  2005.}}

\author{M. D. Maia}
\address{Instituto  de F\'\i sica, Universidade de Brasilia,
 70919-970  Brasília  D.F.  Brazil
\\ maia@unb.br}

\begin{abstract}
 The correspondence  between conformal covariant fields
  in Minkowski's  space-time and isometric  fields in the
  five dimensional  anti-deSitter  space-time is   extended to
   a six-dimensional bulk space  and its regular  sub-manifolds,
    so  as   to  include the analysis  of  evaporating
     Schwarzschild's  black holes  without  loss  of
      quantum unitarity.
\end{abstract}

\maketitle

\section{Introduction}
In  1975    S.  Hawking  presented  his  well known
 information  loss theorem     based on  the  semiclassical
  Einstein's equations:
\[
R_{\mu\nu}-\frac{1}R g_{\mu\nu}  =  8\pi G  <T_{\mu\nu}>
\]
as  applied to the  ordinary  Kerr and  Schwarzschild  black holes
 in  four-dimensional space-times. Three possible outcomes were
 predicted:  Either  the
  black hole  evaporates leaving no trace of the  properties   of
   particles  falling  inside it  or,  after the evaporation
   there is  a naked  singularity   or,   lastly, there  would  be a  regular  remnant  of the   black hole.
This result motivated  a 30 year   debate  about the  validity
 of  the quantum unitarity   near  a black hole,
 leading to the conclusion that either  something was missing
 in the theorem
 or  that  quantum  theory  in the regime  of  strong gravity
  should be modified \cite{Hawking:1}.

In  2004   Hawking presented  a new  version of the
  theorem with new  and   different   hypothesis:  Now
   quantum gravity is  approached by  Euclidean Path Integral
    followed by a Wick rotation.  Instead of the usual black holes
    the theorem refers to
extremal black holes in the  $AdS_5$  bulk. Finally it   makes
 use of  the  ADS/CFT correspondence.
The  conclusion is  also different: The  quantum  unitarity is
preserved near the  extremal  black hole  and    some information
 can be recovered  \cite{Hawking:2}.

The  use  of  different  assumptions  makes it  difficult
 to  compare   the   two  theorems on equal   footing.  If in 1975
  ordinary  Special Relativity  with the Poincar\'e
   symmetry  was used, now  the conformal  symmetry  of Minkowski's  space is  used, leaving  the impression is that  we  are playing a different game. Nonetheless, when discussing the  subject in classrooms we  can hardly  avoid questions  such as:  but  then,  what happened  with the    good  old  Schwarzschild's  solution?  Are  the Schwarzschild  black  holes  still around?  If so,  can we still apply  the  ADS/CFT  correspondence? The purpose of this talk  is  to show that it is possible to extend  that  correspondence in such  way that  the  quantum  unitarity near  a   Schwarzschild  black hole can also be   ensured.

\section{Conformal Symmetry}

Back in  1909,  just  after   the  definition of Minkowski's
 space-time $M_4$  whose  metric is  invariant  under the
  Poincar\'e
  symmetry,   it was found that   electrodynamics  was  also
   covariant under a  larger   symmetry, the   conformal group
     defined by Minkowski's metric  \cite{Cunningham}.
     As  it was later  shown by I. Segal,   this   is in fact the
     most general  symmetry admitted by  Maxwell's
      equations \cite{Segal}. Therefore,  following the  same
       reasoning  of  Minkowski, we  may ask  if there  would be
        a new  \emph{Conformal Special Relativity,  with a new
         space-time    such  that its  metric is  invariant under
         the  conformal group of  $M_4$}.

Even before  attempting an  answer,  it was  found that
the use of   conformal symmetry  in electrodynamics  was  hampered
by  the   causality principle. For  example, to  maintain the
conformal  covariance the solutions  of the electromagnetic
potential wave  equations must  include the
advanced $A_{\mu}(x+vt)$ and retarded  $A_{\mu}(x-vt)$
 components  together. The presence of  the advanced component
  implies in a violation of the  causality principle,   which
    was then and still is  today   a  principle based on a solid
    intuition: \emph{The past must be  divided from the future by
    the present.  A  denial of these facts would be  a denial of
    our most primitive  intuitions  about  time-order}  \cite{Synge}.

On the other hand,  the   discovery of the four-dimensional
anti-deSitter ($AdS_4$) solutions of  Einstein's  equations
 in 1917  has  shown  that causality may also be  violated  by
 gravity, essentially     because the anti-deSitter  solution
  admits    closed  time-like  geodesics. Thus, if  Einstein once
  gave us  a  physical meaning  to the Riemann Geometry
of abstract  manifolds,  deSitter  demonstrated that  such meaning
 is not necessarily intuitive.

In spite of this  non causal  implication, it was found much later
 that  the  Lie  algebra  of  the   anti-deSitter  group is  more
  consistent  with  the  super-symmetry  than  the  Poincar\'e group
   \cite{Wit}.  This   means  that  at least in some theoretical
    situations   we  may be induced to trade causality  violations
     by some elaborated  formal  development,  even if that theory
      has not yet been proven experimentally. A powerful  example of
      this is given by  the realization of  type II  string theory
      with $E_8\times E_8$  in  the  space  $AdS_5 \times  S^5$
      \cite{HW}.

The  interest  in  conformal symmetry  reappeared in   1967,
when  R. Penrose pointed out that  the semi-direct product structure
 of the Poincar\'e group:
 $P_4 = SO(3,1) \, \bigodot\!\!\!\!\! s  \,\,  T_4$
  implies that the  non-homogeneous  action of the  translation
   subgroup  $T_4$ has to be  handled  separately from the
    homogeneous subgroup in  the   spinor representations
    of $P_4$.   He suggested that the use of the   conformal group
    leads to a  more consistent  spinor representation
    \cite{Penrose:1}.  The resulting spinor  structure,  called
     twistors,   is well defined  in  Minkowski's  space-time  $M_4$,  but  it still resists    the generalization  to  curved   space-times,  solutions of  Einstein's  equations.

It should be  recalled  that  a  stronger  objection  to the
translational subgroup $T_4$   existed  at that  time, stating that
in  any attempt to   combine the   Poincar\'e group with an internal
 or  gauge  symmetry,  all particles  belonging to the same  spin
  multiplet  would   necessarily  have  zero squared mass
  differences. This is  obviously  against the  experimental
   evidences. After much debate  it was  found that  the  problem
    lies within the    subgroup $T_4$,  which  acts   nilpotently
      over  the remaining of the combined symmetry
      \cite{O'Raifeartaigh,Mandula}.  As we know  today, the \emph{adopted solution} of  such No-Go problem as  it  become to be known,  resides  in   the   hypothesis  that  the  combined symmetry  is spontaneously broken by  the Higgs mechanism.
Nonetheless, in the present context,  it  is relevant   to stress
 that if the Poincar\'e   symmetry is  replaced by the  deSitter or
  the anti-deSitter groups, or in fact  by any  semi-simple group,
   then  the No-Go problem would  be solved  exactly, without
   supersymmetry, giving  an additional support  to Penrose's
   argument  in favor of  the  conformal symmetry  \cite{Roman,Maia}.

The next step  in the development of the conformal symmetry in
  relativist  physics  was the  invention of the  ADS/CFT
   correspondence  in  1998 \cite{Maldacena:1}. This is in
   essence  a  relation  between  conform-invariant  fields
   in $M_4$ and the  five-dimensional anti-deSitter gravitational
   field $AdS_5$  \cite{Maldacena:2}. Although  the correspondence
    was  established in the realm of  M-theory, or
     more  specifically  in the properties of  string theory in
     the   $AdS_5 \times  S^5$  space,  it  can  be  applied  to
      most   conform-covariant  fields.

To understand the basics of this correspondence, recall that  the
 four-dimensional anti-deSitter space-time  $AdS_4$ can be
  described as a hypersurface of negative constant  curvature
  isometrically  embedded in the flat space $M_5 (3,2)$.
  By simply adding   one  extra   space-like   dimension, we
  can easily see that  the five-dimensional negative constant
  curvature  surface   $AdS_5$ can also be seen  as  a
   hypersurface embedded    in the flat space $M_6 (4,2)$.
   Since $AdS_5$ is  a maximally symmetric  sub-space  of
   $M_6 (4,2)$,  it follows that any  (pseudo)  rotation  on  that
     sub-space  corresponds  to  a conformal transformation
      in $M_4$,   in accordance with the isomorphism
$$M_4 \stackrel{conformal}{\Longleftarrow} C_o
  \sim S0(4,2) \stackrel{isometric}{\Longrightarrow}  AdS_5$$
where it was  emphasized that while $C_o$ acts conformally on
 $M_4$,   $SO(4,2)$ acts  isometrically on  $AdS_5$.
Therefore,  for  a given   conformally invariant
  field defined in $M_4$,  there  should be  a corresponding
isometrically invariant field  in   the gravitational field of the
 $AdS_5$ space.

\underline{\emph{Example 1:  Twistors on the Anti-deSitter space}}\\
Since  twistors  are  elements of the  spinor  representation of
the    conformal group   $C_o$  on  $M_4$, the above  isomorphism
implies that they   can  also be seen  as the  spinor
 representations of the  group $SO(4,2)$,  here  restricted to be
 the  group of  isometries of the  $AdS_5$  space \cite{Murai}.
  Therefore, the   $M_4$-conformal/$AdS_5$-isometric
  correspondence  defines  twistors   in the  five-dimensional
   gravitational field  of  $AdS_5$.

\underline{\emph{Example 2: Gauge Fields on the $AdS_5$}}

The super-symmetric  Yang-Mills  gauge fields in  $AdS_5$ can be
consistently defined   in the   heterotic
$E_8 \times  E_8$ string theory on the  10-dimensional  space
$AdS_5 \times  S^5$. Therefore, using the ADS/CFT
correspondence, in the reverse order  we may derive  a
 superconformal Yang-Mills field in  $M_4$, corresponding to
 the isometric  Yang-Mills field  in  $AdS_5$.  Reciprocally, the
  superconformal Yang-Mills field in  $M_4$  corresponds  to a
 quantum gauge  field    defined on  the  gravitational environment
   of  the $AdS_5$,  preserving  its main  quantum properties,
   including  the unitarity.

\section{The $M_6 (4,2)$-isometric/$M_4$-conformal correspondence }
The  above  correspondence should  be  somehow   consistent with
 four-dimensional physics and  not  just the  $AdS_5$. This  may
  be  achieved  by  use of the  brane-world theory where, like in
  the popular Randall-Sundrum model the  space $AdS_5$ represents
  the bulk.   According to this  scheme,  all  gauge fields  are
    confined to  the four-dimensional
brane-worlds, but the  geometry  of these subspaces
 propagates along  the extra  dimension.  The  latter condition
 requires that   all perturbations of the brane-world  geometry
 should  belong to the class of  four-manifolds  embedded in the
 same  $AdS_5$  bulk. In order to  see this  we  require some
  basic   understanding of the geometry of  subspaces.

For  simplicity we  consider  here only the case of    the
 4-dimensional brane-world $V_4$   embedded in  a  five dimensional
 bulk space  $V_5$,  as  given by the embedding map
  ${\cal Z}^A : V_4  \rightarrow   V_5$
  (Here  the indices  A,B,C...run from  1 to  5. The indices
   $\mu$,$\nu$..., run from 1  to  4). Together  with the unit
   normal vectors  $\eta^A$, they define a   5-bein
   $\{{\cal Z}^A_{,\alpha},\eta^B\}$,
   in which the  components  of the bulk's  Riemann  tensor
    $^5{\cal R}_{ABCD}$  can be expressed  in terms of the
    brane-world  metric $g_{\mu\nu}$ and the   extrinsic  curvature
     $k_{\mu\nu}$  as
\[
\left\{
\begin{array}{ll}
^5{\cal R}_{ABCD}{\cal Z}^{A}_{,\alpha}
{\cal Z}^{B}_{,\beta}{\cal Z}^{C}_{,\gamma}{\cal Z}^{D}_{,\delta} =
R_{\alpha\beta\gamma\delta}\! -\! 2k_{\alpha[\gamma
}k_{\beta]\delta }\; \;\; \mbox{(Gauss)}\vspace{2mm}\cr
^5{\cal R}_{ABCD}
{\cal Z}^{A}_{,\alpha} \eta^{B}{\cal Z}^{C}_{,\gamma}{\cal
Z}^{D}_{,\delta} = k_{\alpha[\gamma; \delta]}\hspace{2.2cm}
 \mbox{(Codazzi)}
\end{array}
\right.
\]
These  expressions become  the Gauss-Codazzi  equations  when
the Riemannian  metric of the bulk ${\cal G}_{AB}$ is  given. In
 the present  case  this geometry comes from the Einstein-Hilbert
  principle. Starting from this principle,   the dynamics  for the
    embedded  brane-world can  be  derived  from the above
    equations. Indeed,  from  Gauss'  equations  we obtain
\begin{eqnarray*}
 ^5{\cal  R}_{AB}Z^{A}_{,\mu}Z^{B}_{,\nu} = R_{\mu\nu}\!\!  &&-  (g^{\alpha\beta}k_{\mu\alpha } k_{\nu \beta }\!\! -\!\! h k_{\mu\nu})\\
&&+ ^5{\cal R}_{ABCD}\eta^{A} Z^{B}_{,\mu}Z^{C}_{,\nu}\eta^{D}
\end{eqnarray*}
\[
^5{\cal R}  =R  - (K^{2} -h^{2})  + 2\;\;  ^5{\cal R}_{AB} \eta^{A}\eta^{B}
\]
where we have denoted $K^{2}=k^{\mu\nu}{}k_{\mu\nu}\;\;\; h=g^{\mu\nu}k_{\mu\nu}$.
 Therefore, the  Einstein-Hilbert  action for the bulk decomposes
 into  the  brane-world  geometry as
\begin{eqnarray*}
\int \!\! ^5{\cal R}\sqrt{ -{\cal G}}d^{5}v\!\!\equiv &&
 \!\int \!\!\left\{ R\!\! -\!\!(
  K^{2}\!\! - \!\!h^{2})\!+\!2\;
 ^5{\cal R}_{AB}\eta^{A}\eta^{B} \right\}\!\!\sqrt{-\cal{ G}}
 d^{5}v\\  &&=\alpha_{*}\!\!\int\!\! {\cal L}^{*}\sqrt{ -{\cal G}}
 d^{5}v
\end{eqnarray*}
where  we have  included  the Lagrangian ${\cal L}^*$ for the
confined matter  and  gauge  fields.

Now, in the cases of constant  curvature  bulks, like in the $AdS_5$ case,   we have
\[
^5{\cal R}_{ABCD}= \frac{\Lambda_*}{6}({\cal G}_{AC}{\cal G}_{BD}-{\cal G}_{AD}{\cal G}_{BC})
\]
where  $\Lambda_*$ is  a bulk cosmological constant.  Replacing in
the above equations,  and applying to a  spherically symmetric
brane-world, it follows that  the extrinsic  curvature  has   a
particular form $k_{\mu\nu} =\alpha_{0} g_{\mu\nu}$, where $\alpha_0$
is   an integration  constant  which  cannot  be  zero,  under the
penalty of producing just a   trivial (a plane) solution
$k_{\mu\nu}=0$. Therefore, the Schwarzschild's solution embedded in
$AdS_5$ necessarily becomes  a Schwarzschild-deSitter  space:
\begin{eqnarray*}
ds^2\!\! =\!\! (1\!\! -
\! \frac{2m}{r}\!+\beta_0^2 r^2)^{-1}dr^2  +r^2 d\omega^2
-(1\!\!-\!\!\frac{2m}{r} + \beta_0^2 r^{2})dt^2
\end{eqnarray*}
where  $\beta_0^2 =( 3\alpha_{0}^2 -\Lambda_* )$. This  shows that
not all four-manifolds can be embedded in the   $AdS_5$ bulk without
imposing  a constraint in its extrinsic geometry. In particular the
Schwarzschild black hole cannot fit into the  $AdS_5$ bulk.
\vspace{2mm}\\ The  above arguments show   why the Schwarzschild
black holes  were not considered  in \cite{Hawking:2}. On the other
hand, it is not  clear to us that extremal black-holes can be defined
and be perturbatively  stable in the sense of Nash's theorem
\cite{Nash} within the $AdS_5$ bulk. In order to reinstate the
Schwarzschild black hole  we notice  that the spaces $AdS_5$ and $M_6
(4,2)$ have the same 15-parameter group of isometries $SO(4,2)$, such
that all arguments based  on the symmetries  of the   $AdS_5$, can be
extended to $M_6 (4,2)$.  In addition, it can be   extended to all
four-dimensional isometrically embedded submanifolds  of the $M_6
(4,2)$ bulk  and not just those  which are embedded in its $AdS_5$
hypersurface. With this in mind  we propose an extension of the
ADS/CFT correspondence to this larger  class of four-dimensional
subspaces of the $M_6 (4,2)$, provided  we may keep a   1:1
correspondence \cite{Maia:2}. A local realization of this  extension
can be obtained by use of the inverse functions theorem,  when the
embedding functions  are  also regular. This  is precisely  the
condition which required by Nash's theorem applied to find
differentiable solutions of the Gauss-Codazzi equations. \emph{The
combination of the  $M_6 (4,2)$-isometric/ $M_4$-conformal
correspondence with   the regular and differentiable embeddings
provides  an extended form of conformal to isometric correspondence},
whereby  a  conformally invariant field in $M_4$ corresponds to field
on  an isometrically and regular embedded brane-world field in $M_6
(4,2)$. As it happens, $M_6 (4,2)$ is the regular embedding space for
the Schwarzschild solution,  so that using such extension the
unitarity  of the quantum gauge fields near a Schwarzschild  black
hole may be implemented.

\end{document}